\documentclass{iaus}
\usepackage{graphics}

  \checkfont{eurm10}
  \iffontfound
    \IfFileExists{upmath.sty}
      {\typeout{^^JFound AMS Euler Roman fonts on the system,
                   using the 'upmath' package.^^J}%
       \usepackage{upmath}}
      {\typeout{^^JFound AMS Euler Roman fonts on the system, but you
                   dont seem to have the}%
       \typeout{'upmath' package installed. iaus.cls can take advantage
                 of these fonts,^^Jif you use 'upmath' package.^^J}%
      }
  \else
  \fi

  \checkfont{msam10}
  \iffontfound
    \IfFileExists{amssymb.sty}
      {\typeout{^^JFound AMS Symbol fonts on the system, using the
                'amssymb' package.^^J}%
       \usepackage{amssymb}%
         
       \let\ge=\geqslant  
      }{}
  \fi

  \IfFileExists{amsbsy.sty}
    {\typeout{^^JFound the 'amsbsy' package on the system, using it.^^J}%
     \usepackage{amsbsy}}
    {\providecommand\boldsymbol[1]{\mbox{\boldmath $##1$}}}

\newsavebox{\astrutbox}
\sbox{\astrutbox}{\rule[-5pt]{0pt}{20pt}}

\title[Outskirts of Galaxy Clusters: intense life in the suburbs]
      {A Detailed View of the Fundamental Plane of Early--Type Galaxies in
       Clusters at $z\sim0.2$}

\author[A. Fritz \& B.~L.\ Ziegler]{Alexander Fritz$^1$ \and Bodo L.\ Ziegler$^1$}

\affiliation{$^1$Universit\"ats--Sternwarte G\"ottingen, Geismarlandstra{\ss}e 11,
37083 G\"ottingen, Germany email: afritz@uni-sw.gwdg.de\\[\affilskip]}

\pubyear{2004}
\volume{195}
\pagerange{1--8}
\date{?? and in revised form ??}
\jname{Outskirts of Galaxy Clusters: intense life in the suburbs}
\editors{A. Diaferio, ed.}
\begin{document}

\maketitle

\begin{abstract}
We present a spectroscopic sample of 48 early-type galaxies in the rich
cluster Abell~2218 and 50 galaxies in Abell~2390. Since both samples are
very similar, we combine them and investigate a total number of 98
early-type galaxies at $z\sim 0.2$.
A subsample of 34 galaxies with \textit{HST} structural properties
is used to construct the Fundamental Plane. Elliptical and S0 galaxies show
a zeropoint offset of $\overline{m}_{r}\sim0.43$~mag with respect
to the local Coma FP. Both sub-samples, ellipticals and lenticulars,
exhibit a similar, mild evolution and small scatter.
The moderate amount of luminosity evolution is consistent with stellar
population models of passive evolution, if $z_{f}\ge2$ is assumed.
\end{abstract}

\firstsection 
\section{Introduction}
 
Clusters of galaxies are powerful laboratories in investigating the formation
and evolution of galaxies. Rich clusters are dominated by early-type (E+S0)
galaxies. A detailed study of the properties of these galaxies provides
beneficial insights in their formation and evolution. In particular, the
paradigm of hierarchical galaxy formation can be critically tested.

In the local Universe numerous investigations revealed that E+S0 form a very
homogeneous galaxy population (e.g., \cite{BLE92}; \cite{BBF93}).
Their structural parameters (effective surface brightness $\langle\mu\rangle_e$
and the size as described by the effective radius $R_e$) and kinematics
(velocity dispersion $\sigma$) represent a tight correlation in three
dimensional parameter space, the Fundamental Plane (\cite{DD87}).
Furthermore, they show a small scatter in their relations of colours
(e.g., Mg--($B-V$)), $M/L$ ratios and absorption line indices with velocity
dispersion (e.g., Mg--$\sigma$). However, the question arises whether E+S0s are
truly one single family or rather a diverse group with different formation and
evolutionary processes. Therefore, one aim of this work is to explore if there
are differences between the properties of elliptical and S0 galaxies.

Previous spectroscopic studies were limited to a small number of the more
luminous galaxies. To overcome bias and selection problems of small samples,
we focus in this study of the clusters Abell~2218 (\cite{Z2001}) and Abell~2390
on a large number of objects ($N=98$), spanning a wide range in luminosity, in
case of A\,2390 $21.4<B<23.3$, and a wide field-of-view (FOV) of
$\sim 10'\times 10'$ ($1.56\ h^{-1}_{70}\times 1.56\ h^{-1}_{70}$~Mpc$^{2}$).
We adopt a cosmology for a flat Universe with $\Omega_{m}=0.3$,
$\Omega_{\Lambda}=0.7$ and $H_0=70$\,km\,s$^{-1}$\,Mpc$^{-1}$.

The objects were selected on the basis of ground-based Gunn $i$-band images
and a combination of defined colour regions using a similar selection procedure
as described in Ziegler et al. (2001). In particular, in the selection
procedure for the mask design special care was taken that galaxies cover the
whole FOV. Additional imaging data obtained with the 200~inch Hale telescope 
is available in $U$ and $B$ and \textit{HST} WFPC2 observations in the F555W
and F814W filter.

\section{Observations and Data Reduction for Abell~2390}

During two observing runs (09/1999 and 07/2000) we observed three masks using
the MOSCA spectrograph at the Calar Alto 3.5~m telescope with total exposure
times between $\sim$8 and $\sim$12 hours each. In total, we obtained 63 high
$S/N$ spectra of 50 different early-type galaxies, out of which 15 are situated
within the \textit{HST} field. Only one object was revealed to be a background
galaxy, which shows that our sample selection was highly efficient.
The instrumental resolution in the spectral regime of ${\rm H}\beta$ and
${\rm Mg}\, b$ was 5.5~\AA\ FWHM ($\sigma_{\rm inst}\sim 100$~km~s$^{-1}$). 
The average $S/N$ is $\sim 41$.

The reduction of the spectra was carried out using standard reduction
techniques and velocity dispersions were calculated with the FCQ method
(\cite{Ben:90}). Structural parameters were determined by fitting the surface
brightness profile with either an $r^{1/4}$ or a combination of an $r^{1/4}$
plus exponential law profile (\cite{Sag:97a}). Further details concerning the
reduction and analysis are outlined in Fritz et al. (2004).

\begin{figure}
\begin{center}
\vspace{0.5cm}
\resizebox{7.0cm}{!}{\includegraphics{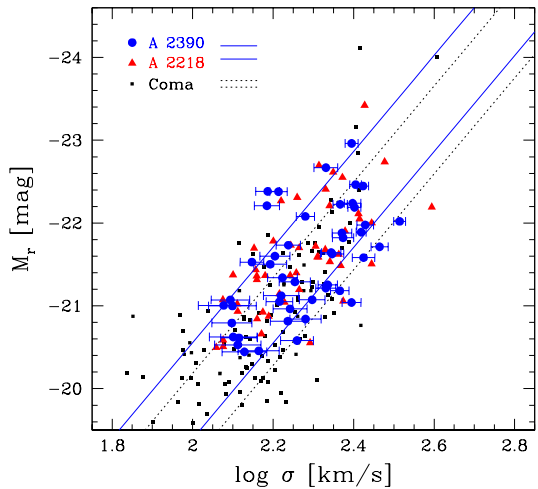}}
\caption{Gunn $r$ Faber-Jackson relation for 94 early-type cluster galaxies in
  A\,2390 and A\,2218, compared to the Coma sample (JFK95). The Coma galaxies
  were restricted to $M_{r}<-19.5$ and log~$\sigma>2.0$. The area
  bounded by dotted lines indicates the mean $\pm1~\sigma$ of the local FJ
  relation using a linear bisector fit, solid lines the resp. fits to the
  distant sample within $\pm1~\sigma$.}
\label{fig:FJ}
\end{center}
\end{figure}

\section{A detailed Fundamental Plane of E+S0 Cluster Galaxies at $z\sim0.2$}

\begin{figure}
\begin{center}
\vspace{0.5cm}
\resizebox{12.5cm}{!}{\includegraphics{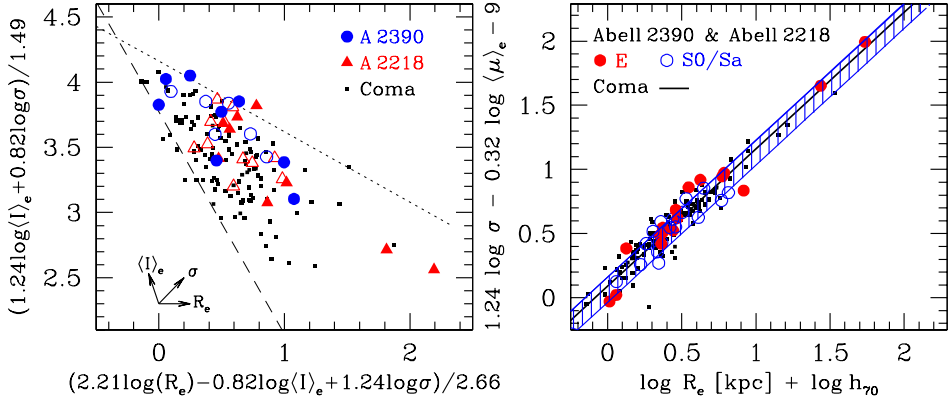}}
\caption{Fundamental Plane for A\,2390 and A\,2218 in Gunn $r$. {\it Left}:
  Face-on view. The dotted line indicates the exclusion zone for nearby
  galaxies, the dashed line the $L$ limit for Coma. Filled symbols represent
  ellipticals, open symbols S0 galaxies. {\it Right}: Edge-on FP for the
  combined cluster sample, with a separation into morphologies. The solid line
  marks the Coma relation of JFK95. Fits along 1~$\sigma$ errors for the S0
  galaxies assuming a constant slope are indicated (hashed area).}
\label{fig:FP}
\end{center}
\end{figure}

In Fig.\,\ref{fig:FJ} the Faber-Jackson (FJ) relation for the complete sample of
94 E+S0 galaxies is shown. As a local reference we use the Coma sample of
\cite{JFK95} (JFK95). Overall, the luminosity evolution is small
($\overline{m}_{r}\sim0.32$~mag). The Fundamental Plane (FP) in rest frame Gunn
$r$ is shown in Fig.\,\ref{fig:FP}. The sub-sample with \textit{HST} structural
parameters comprises 34 E+S0 galaxies, splitted into 17 ellipticals, 2 E/S0,
8 S0, 3 SB0/a, 3 Sa and 1 Sab galaxy that enter the FP. With this large sample,
possible differences of the galaxies' properties can be explored for various
sub-populations. Both clusters have a similar behaviour within and along the FP,
with no hint for a slope change. For the 34 cluster E+S0 galaxies we find a
combined evolution of $\overline{m}_{r}\sim0.43$~mag. Lenticular galaxies 
(E/S0, S0, SB0/a, Sa, Sab) have a zeropoint offset of $0.30$~mag with
respect to Coma and seem to inhabit a certain band within the FP
(Fig.\,\ref{fig:FP}, right). Ellipticals exhibit a similar mild evolution of
$\overline{m}_{r}\sim0.46$. Both sub-samples have a small scatter
($\sim$0.10 mag).

\section{Summary}

We construct a spectroscopic sample of 48 E+S0 in the rich cluster
Abell~2218 ($z=0.18$) and 50 E+S0 in Abell~2390 ($z=0.23$). Since both cluster
samples are very similar, we combine them and analyse a total number of
98 E+S0 at $z\sim 0.2$. For a subsample of 34 galaxies we explore the evolution
of the FP and study their $M/L$ ratios. On average, all 34 cluster
E+S0 galaxies show a mild evolution of $\Delta {\rm log} (M/L_{r})=-0.17$.
S0 galaxies show lower $M/L$ ratios of
$\Delta {\rm log} (M/L_{r})_{\rm S0}=-0.12$ with respect to the local reference.
For the ellipticals we again deduce a moderate evolution of
$\Delta {\rm log} (M/L_{r})_{\rm E}=-0.18$. Elliptical and S0 galaxies seem
to comprise similar sub-populations with no significant differences. An analysis
based on the $M/L$ ratios revealed a mean formation redshift of $z_{f}\sim3$,
which is consistent with stellar population models of passive evolution.

\begin{acknowledgments}
This project is a collaboration with R.~G.\ Bower and I. Smail (Durham/UK)
and R.~L.\ Davies (Oxford/UK) who have contributed to these results. We
thank the Calar Alto staff for efficient observational support. AF and BLZ
acknowledge financial support by the Volkswagen Foundation (I/76\,520) and
the DFG (ZI\,663/1-1, ZI\,663/2-1).
\end{acknowledgments}

\end{document}